\documentclass{aa}

\usepackage{graphicx}
\usepackage{epsfig}

\newcommand{\pl}{\partial}
\renewcommand{\d}{{\rm d}}

\newcommand{\beq}{\begin{equation}}
\newcommand{\eeq}{\end{equation}}
\newcommand{\beqa}{\begin{eqnarray}}
\newcommand{\eeqa}{\end{eqnarray}}
\newcommand{\bea}{\begin{array}}
\newcommand{\ea}{\end{array}}
\newcommand{\bx}{{\bf x}}

\newcommand{\bq}{{\bf q}}

\newcommand{\bk}{{\bf k}}
\newcommand{\lag}{\langle}
\newcommand{\rag}{\rangle}
\newcommand{\bv}{{\bf v}}
\newcommand{\Om}{\Omega_{\rm m}}

\newcommand{\dR}{\delta_{R}}
\newcommand{\dL}{\delta_L}
\newcommand{\DL}{\Delta_L}

\newcommand{\xib}{\overline{\xi}}
\newcommand{\Det}{{\rm Det}}
\newcommand{\Tr}{{\rm Tr}}

\newcommand{\psib}{\overline{\psi}}

\newcommand{\ad}{{\dot a}}
\newcommand{\cO}{{\cal O}}

\newcommand{\dy}{\delta_y}
\newcommand{\dRG}{\delta_R^G}
\newcommand{\cP}{{\cal P}}

\begin{document}

% US additions
 
%\topmargin=3 cm
%\evensidemargin=2.5 cm
%\oddsidemargin=2.5 cm
%\thispagestyle{empty}

\title{Dynamics of gravitational clustering V. Subleading corrections in the quasi-linear regime.}   
\author{P. Valageas}  
\institute{Service de Physique Th\'eorique, CEN Saclay, 91191 Gif-sur-Yvette, France} 
\date{Received / Accepted }

\abstract{
We investigate the properties of the standard perturbative expansions which describe the early stages of the dynamics of gravitational clustering. We show that for hierarchical scenarios with no small-scale cutoff perturbation theory always breaks down beyond a finite order $q_+$. Besides, the degree of divergence increases with the order of the perturbative terms so that renormalization procedures cannot be applied. Nevertheless, we explain that despite the divergence of these subleading terms the results of perturbation theory are correct at leading order because they can be recovered through a steepest-descent method which does not use such perturbative expansions. Finally, we investigate the simpler cases of the Zel'dovich and Burgers dynamics. In particular, we show that the standard Burgers equation exhibits similar properties. This analogy suggests that the results of the standard perturbative expansions are valid up to the order $q_+$ (i.e. until they are finite). Moreover, the first ``non-regular'' term of a large-scale expansion of the two-point correlation function should be of the form $R^{-2} \sigma^2(R)$. At higher orders the large-scale expansion should no longer be over powers of $\sigma^2$ but over a different combination of powers of $1/R$. However, its calculation requires new non-perturbative methods.
}

\maketitle

\section{Introduction}

In standard cosmological scenarios large-scale structures in the universe have formed through the growth of small initial density perturbations by gravitational instability, see \cite{Peebles1}. Besides, in most cases of cosmological interest the power increases at small scales as in the CDM model (\cite{Peebles3}). This leads to a hierarchical scenario of structure formation where smaller scales become non-linear first. Therefore, one often uses a perturbative approach in order to describe the density field at large scales or early times. This is of practical interest in order to study for instance weak lensing effects on large scales, which may in turn constrain the cosmological parameters (e.g., \cite{Ber3}). The standard perturbative expansions are derived from the equations of hydrodynamics, coupled with Poisson equation, by writing the density field as a perturbative expansion over the initial density contrast (or rather over the linear growing mode). Such a procedure is detailed in \cite{Fry1} and \cite{Goroff1}. As noticed in a previous work (\cite{paper1}) one obtains the same perturbative expansion from the collisionless Boltzmann equation. This clearly shows that perturbation theory cannot handle shell-crossing (in fact, it does not capture any effect arising from multi-streaming) and it only provides asymptotic expansions. Moreover, as noticed in \cite{Lok1} or \cite{Scoc1} the terms encountered in perturbation theory actually diverge at next-to-leading order for steep linear power-spectra $P_L(k) \propto k^n$ with $n \geq -1$. This shows that perturbation theory is rather ill-defined and it casts some doubts on the validity of its predictions, even when they are finite.

Thus, in this article we reconsider the properties of the standard perturbative expansions which describe the dynamics of gravitational clustering. First, we recall in Sect.\ref{Standard perturbative expansion} the derivation of these perturbative expansions from the equations of motion. Then, we show in Sect.\ref{Small scale divergences} that for hierarchical scenarios where the power increases at small scales with no bound (e.g., linear power-spectra with $n>-3$) perturbation theory always breaks down beyond a finite order $q_+$ as divergent terms appear. However, we explain in Sect.\ref{Steepest-descent method} that, despite these divergences, the results of perturbation theory for the cumulants of the density contrast are correct at leading-order because they can also be obtained from a steepest-descent method which does not make use of such perturbative expansions. Finally, we consider in Sect.\ref{Comparison with simple dynamics} the simpler cases of the Zel'dovich and Burgers dynamics. In particular, we show that the standard Burgers equation shares many properties with the gravitational dynamics. This suggests that the results of perturbation theory are valid up to the order $q_+$ (i.e. until they are finite) and that beyond this order the large-scale expansion of the two-point correlation (for instance) is no longer over powers of $\sigma^2(R)$ but over some different combination of powers of $1/R$.

\section{Standard perturbative expansions}
\label{Standard perturbative expansion}

\subsection{Perturbative approach}
\label{Equations of motion}

In this section we recall the standard perturbative method which is used to describe the early stages of gravitational dynamics in an expanding universe. For simplicity we focus on a critical-density universe (i.e. $\Om=1$) but our results also apply to other cosmological parameters. Since we are interested in the large-scale behaviour of the density field we adopt as usual an hydrodynamical description. Thus, the matter distribution is approximated as a pressureless fluid with no vorticity. Besides, we only consider the linear growing mode. Then, one can look for a perturbative solution to the equations of motion for the density contrast $\delta(\bx,t)$ and the divergence $\theta(\bx,t) \equiv \nabla . \bv$ of the peculiar velocity $\bv(\bx,t)$ (indeed, within perturbation theory we can take the velocity field to be irrotational). Here we introduced the comoving coordinate $\bx$. As shown in \cite{Goroff1} the time dependence factorizes so that these expansions have the form:
\beq
\delta(\bx,t) = \sum_{l=1}^{\infty} a^l(t) \; \delta^{(l)}(\bx) , \;\;\; \theta(\bx,t) = \sum_{l=1}^{\infty} \ad a^l \; \theta^{(l)}(\bx) ,
\label{exp1}
\eeq
where $a(t) \propto t^{2/3}$ is the scale-factor and $\ad=\d a/\d t$ is its time-derivative. The terms $\delta^{(l)}$ and $\theta^{(l)}$ do not depend on time and they are of order $l$ over the growing linear density field $\delta^{(1)}(\bx) \equiv \dL(\bx)$. Thus, in the following we note $\dL(\bx)$ the linear density field extrapolated up to $z=0$ (i.e. $a=1$). In fact, since time plays no role (because of the factorization) we shall simply investigate the properties of the density contrast $\delta(\bx)$ today. The quasi-linear regime simply corresponds to the limit of small amplitude for the power-spectrum of the linear field $\dL(\bx)$ (as described below this is usually different from the limit of large scale). In order to compute the perturbative expansions (\ref{exp1}) it is convenient to work in Fourier space. Therefore, we define the Fourier transform of the density field as:
\beq
\delta(\bk) = \int \frac{\d\bx}{(2\pi)^3} \; e^{-i \bk.\bx} \; \delta(\bx) ,
\label{Four1}
\eeq
and similarly for all other fields. Then, substituting the expansions (\ref{exp1}) into the equations of motion one obtains (see \cite{Goroff1}):
\beqa
\delta^{(l)}(\bk) & = & \int \d\bq_1 .. \d\bq_l \; \delta_D(\bk - \bq_1 - .. - \bq_l) \; F_l^{(s)}(\bq_1,..,\bq_l) \nonumber \\ & & \times \; \dL(\bq_1) .. \dL(\bq_l)
\label{Fl1}
\eeqa
and:
\beqa
\theta^{(l)}(\bk) & = & - \int \d\bq_1 .. \d\bq_l \; \delta_D(\bk - \bq_1 - .. - \bq_l) \; G_l^{(s)}(\bq_1,..,\bq_l) \nonumber \\ & & \times \; \dL(\bq_1) .. \dL(\bq_l) .
\label{Gl1}
\eeqa
Here $\delta_D$ is Dirac distribution while $F_l^{(s)}$ and $G_l^{(s)}$ are dimensionless homogeneous functions of the wave vectors $(\bq_1,..,\bq_l)$ with degree zero. They are symmetric over permutations of their arguments. These functions obey the recursion relations obtained in \cite{Goroff1} (see also \cite{Jain1}). In addition, it is useful to note that the kernels $F_l^{(s)}(\bq_1,..,\bq_l)$ are proportional to $k^2$ for $l \geq 2$, with $\bk= \bq_1+..+\bq_l$, see \cite{Goroff1}. This is a consequence of momentum conservation (e.g., \cite{Peebles2}, \cite{Hogan1}).

Next, in order to derive the statistical properties of the density field $\delta(\bx)$ we must specify the properties of the linear density field $\dL(\bx)$ over which the perturbative expansions are built. In this article we investigate the case of Gaussian initial conditions and we take the linear growing mode $\dL(\bx)$ to be a random Gaussian field. Therefore, its statistical properties are fully determined by its power-spectrum $P_L(k)$ which we define by:
\beq
\lag \dL(\bk) \dL(\bk') \rag \equiv P_L(k) \; \delta_D(\bk+\bk') .
\label{PL1}
\eeq
Here we explicitly wrote the subscript ``L'' in order to recall that $P_L(k)$ is defined with respect to the field $\dL(\bx)$: it is not the power-spectrum of the actual density contrast $\delta(\bx)$. We shall often consider power-law power-spectra of the form:
\beq
P_L(k) \propto k^n \;\;\; \mbox{with} \;\;\; -3 < n < 1 ,
\label{n1}
\eeq
which cover all cases of cosmological interest. Of course, the actual power-spectrum is not a pure power-law but it is usually sufficiently smooth (like the CDM model) to be approximated by a power-law over the range of interest.

\subsection{The non-linear power-spectrum}
\label{The non-linear power-spectrum}

\begin{figure}[htb]
\centerline{\epsfxsize=8cm \epsfysize=3.9cm \epsfbox{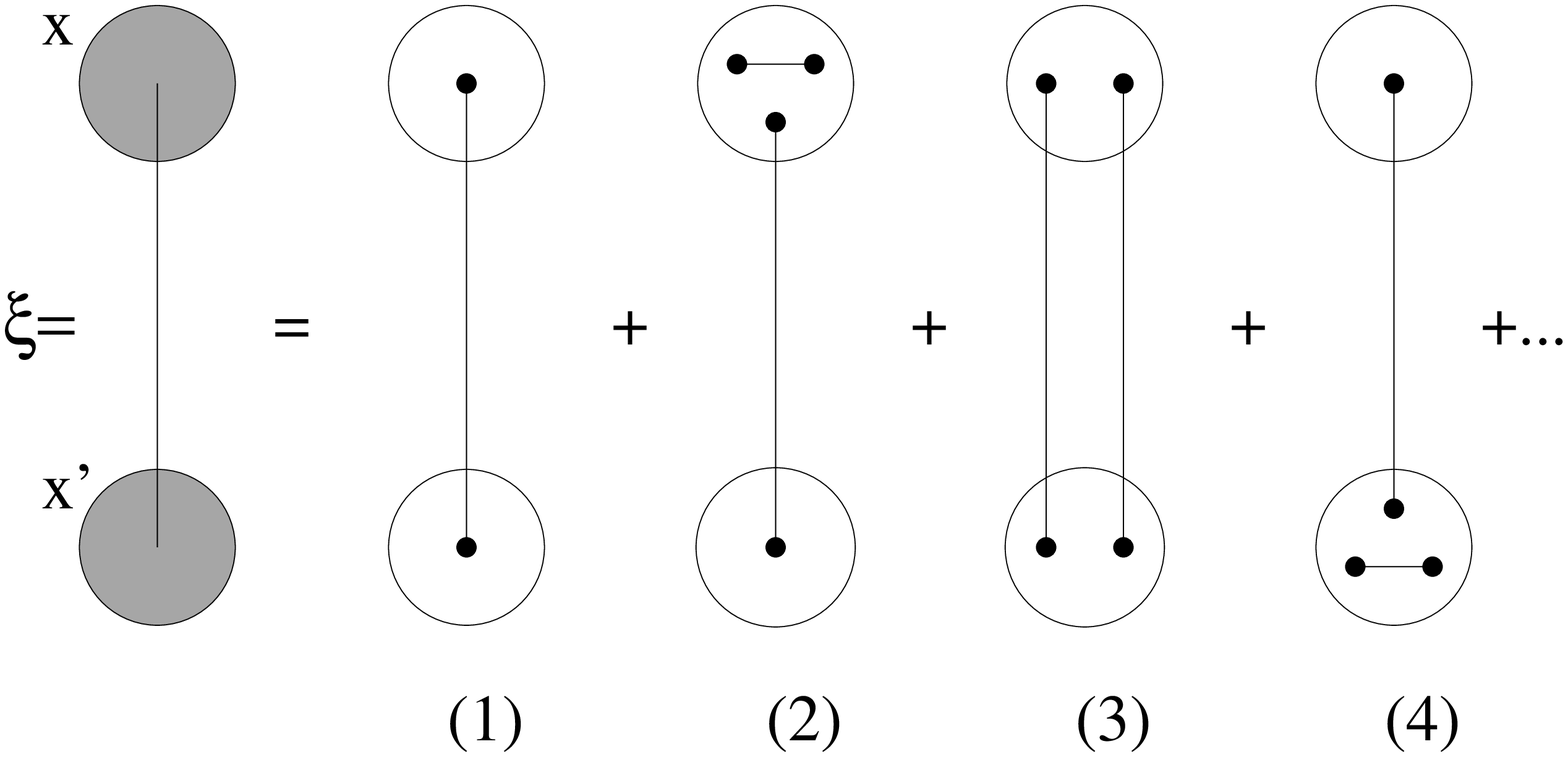}}
\caption{Diagrammatic representation of the standard expansion (\ref{xi2}) for the two-point correlation $\xi(\bx,\bx') \equiv \lag\delta(\bx) \delta(\bx')\rag$. The filled circles on the left represent the actual density contrasts $\delta(\bx)$ and $\delta(\bx')$. A white circle on the right with $l$ points corresponds to $\delta^{(l)}(\bx)$ and each dot represents a factor $\dL$ in the expansion of $\delta^{(l)}$. Each line stands for a factor $\DL$ arising from the Gaussian average. We did not write multiplicity factors.} 
\label{figexp1}
\end{figure}

From the standard perturbative approach recalled in Sect.\ref{Equations of motion} it is possible (at least in principle) to compute the various moments of the actual density contrast $\delta(\bx)$ up to any order $l$ over the field $\dL(\bx)$. In particular, some previous works (e.g., \cite{Jain1}, \cite{Scoc1}) have studied the non-linear power-spectrum $P(k)$ of the density contrast $\delta(\bx)$, which is defined as in eq.(\ref{PL1}) by:
\beq
\lag \delta(\bk) \delta(\bk') \rag \equiv P(k) \; \delta_D(\bk+\bk') .
\label{P1}
\eeq
It is obtained by substituting the expansion (\ref{exp1}) into eq.(\ref{P1}) which yields up to order $P_L^2$:
\beqa
\lefteqn{ P(k) \; \delta_D(\bk+\bk') = \lag \left( \sum_{l=1}^{\infty} \delta^{(l)}(\bk) \right) \left( \sum_{l'=1}^{\infty} \delta^{(l')}(\bk') \right) \rag } \nonumber \\ & & = \lag \delta^{(1)}(\bk) \delta^{(1)}(\bk') \rag + \lag \delta^{(3)}(\bk) \delta^{(1)}(\bk') \rag \nonumber \\ & & \;\;\; + \lag \delta^{(2)}(\bk) \delta^{(2)}(\bk') \rag + \lag \delta^{(1)}(\bk) \delta^{(3)}(\bk') \rag + \cO(\dL^6) .
\label{P2}
\eeqa
The various averages are computed from eq.(\ref{Fl1}) and eq.(\ref{PL1}) using Wick's theorem. Such a study is detailed for instance in \cite{Scoc1} where the expansion (\ref{P2}) is described as a sum over specific diagrams. Although these works were developed in Fourier space $\bk$ it is convenient for our purposes to work in real space $\bx$, at least in the intermediate steps. Thus, we can consider the expansion of the two-point correlation function $\xi(\bx,\bx')$ defined by:
\beq
\xi(\bx,\bx') \equiv \lag\delta(\bx) \delta(\bx')\rag .
\label{xidef1}
\eeq
As in eq.(\ref{P2}), we can introduce the expansion (\ref{exp1}) through:
\beq
\xi(\bx,\bx') = \lag \left( \sum_{l=1}^{\infty} \delta^{(l)}(\bx) \right) \left( \sum_{l'=1}^{\infty} \delta^{(l')}(\bx') \right) \rag .
\label{xi1}
\eeq
In real space, the expression (\ref{Fl1}) for the terms $\delta^{(l)}(\bx)$ reads:
\beq
\delta^{(l)}(\bx) = \int \d\bx_1 .. \d\bx_l \; F_l^{(s)}(\bx;\bx_1,..,\bx_l) \; \dL(\bx_1) .. \dL(\bx_l)
\label{Flx1}
\eeq
where the symmetric functions $F_l^{(s)}$ are the inverse Fourier transforms of the functions $F_l^{(s)}(\bq_1, .. , \bq_l)$ introduced in eq.(\ref{Fl1}):
\beqa
\lefteqn{ F_l^{(s)}(\bx;\bx_1,..,\bx_l) = \int \d\bk \; \frac{\d\bq_1 .. \d\bq_l}{(2\pi)^{3l}} \; e^{i (\bk.\bx-\bq_1.\bx_1-..-\bq_l.\bx_l)} } \nonumber \\ & & \times \; F_l^{(s)}(\bq_1, .. , \bq_l) \; \delta_D(\bk-\bq_1-..\bq_l) .
\eeqa
The functions $F_l^{(s)}(\bx;\bx_1,..,\bx_l)$ are invariant under global translations and rotations. Of course, since the initial conditions are homogeneous and isotropic the two-point correlation $\xi(\bx,\bx')$ only depends on $|\bx-\bx'|$. The expansion (\ref{xi1}) is equivalent to the expansion (\ref{P2}). In particular, we have:
\beq
\xi(x) = \int \d\bk \; e^{i \bk.\bx} \; P(k) ,
\label{xiP1}
\eeq
where we defined the function of one variable $\xi(x)$ by $\xi(\bx_1,\bx_2)=\xi(|\bx_1-\bx_2|)$. Then, using Wick's theorem the expansion (\ref{xi1}) can be written as a power-series over the two-point correlation $\DL(\bx,\bx')$ of the linear density field $\dL(\bx)$, defined by:
\beq
\lag \dL(\bx) \dL(\bx') \rag \equiv \DL(\bx,\bx') .
\label{DL1}
\eeq
Of course, the kernel $\DL(\bx,\bx')$ only depends on $|\bx-\bx'|$ and the function $\DL(x)$ is the inverse Fourier transform of the linear power-spectrum $P_L(k)$ introduced in eq.(\ref{PL1}), as in eq.(\ref{xiP1}). As in eq.(\ref{P2}), the expansion (\ref{xi1}) reads up to order $\DL^2$:
\beqa
\lefteqn{ \xi(\bx,\bx') = \lag \delta^{(1)}(\bx) \delta^{(1)}(\bx') \rag + \lag \delta^{(3)}(\bx) \delta^{(1)}(\bx') \rag } \nonumber \\ & &  + \lag \delta^{(2)}(\bx) \delta^{(2)}(\bx') \rag + \lag \delta^{(1)}(\bx) \delta^{(3)}(\bx') \rag + \cO(\dL^6) .
\label{xi2}
\eeqa
We give in Fig.\ref{figexp1} a diagrammatic representation of the expansion (\ref{xi2}). Such a geometric representation was already displayed in \cite{Ber1}.

Thus, the standard expansion (\ref{xi2}) yields the two-point correlation $\xi(\bx,\bx')$ as a perturbative expansion over the Gaussian random field $\dL$. This can be written as:
\beq
\xi(\bx,\bx') = \sum_{l=1}^{\infty} \xi^{(l)}(\bx,\bx')
\label{xilexp1}
\eeq
where the term $\xi^{(l)}$ is of order $\DL^{l}$. In particular, the lowest-order term (the ``tree-level'' contribution) is simply:
\beq
\xi^{(1)}(\bx,\bx') = \DL(\bx,\bx') .
\label{xil0}
\eeq
It corresponds to the diagram labeled (1) in Fig.\ref{figexp1}. The next-to-leading term $\xi^{(2)}$ is given by the diagrams (2) to (4) in Fig.\ref{figexp1}, with appropriate multiplicity factors. Its explicit expression is easily derived from eq.(\ref{xi2}).

\section{Small scale divergences}
\label{Small scale divergences}

The standard expansions built from the method recalled in Sect.\ref{Standard perturbative expansion} have been used to obtain the cumulants of the density contrast at leading-order (``tree-level'') in \cite{Ber1} and \cite{Ber2}. Then, \cite{Lok1} and \cite{Scoc1} studied the next-to-leading term $P^{(2)}(k)$ (one loop correction). They noticed that for power-law linear power-spectra $P_L$ with $n \geq -1$ this subleading term actually diverges at large wavenumbers $k \rightarrow \infty$. This ultraviolet divergence comes from small scales which are non-perturbative and which cannot be described by the hydrodynamic equations of motion. This means that the standard perturbative expansion breaks down for $n \geq -1$. In fact, as we note below one can see that for $n>-3$ there always exists a finite order $l$ beyond which some of the perturbative terms diverge. Therefore, the perturbative expansion is rather ill-defined. This is not surprising since for $n>-3$ sufficiently small scales are far within the non-linear regime, which cannot be reached by perturbative means. Indeed, perturbative approaches cannot go beyond shell-crossing (see for instance the discussion in \cite{paper1}).

The high$-k$ divergences arise from the integrations over the ``small'' branches contained within one of the two circles $\bx$ or $\bx'$ in Fig.\ref{figexp1}. This can be easily understood as follows. By looking at the two-point correlation $\xi(\bx,\bx')$ we select the large scale $|\bx-\bx'|$ for the factors $\DL$ which arise from the vertical branches joining both circles. This provides a window function of scale $|\bx-\bx'|$ which prevents any high$-k$ divergence. By contrast, there is no such constraint for the factors $\DL$ which come from the small branches within one circle. Then, the integration over the wavenumbers $k$ associated with these small branches can lead to high$-k$ divergences since they do not exhibit any window which would cut the contributions from small scales\footnote{We would like to thank F. Bernardeau for discussions on these points.}. This behaviour occurs for the diagrams (2) and (4) shown in Fig.\ref{figexp1} for a steep power-spectrum $n \geq -1$ but it only appears at higher orders for $n<-1$.

Thus, let us consider the contribution of diagrams of the form (2), with only one large branch of size $\sim |\bx-\bx'| \equiv R$, one factor $\dL$ inside the circle $\bx'$ and $2l+1$ factors $\dL$ inside the circle $\bx$ ($l=1$ in the diagram (2) shown in Fig.\ref{figexp1}). In Fourier space, such a diagram yields a contribution $P_{1,2l+1}(k)$ to the non-linear power-spectrum $P(k)$ ($k \sim 1/R$) with:
\beqa
\lefteqn{ P_{1,2l+1}(k) \; \delta_D(\bk+\bk') \equiv \lag \dL(\bk') \delta^{(2l+1)}(\bk) \rag } \nonumber \\ & = & \int \d\bq_1 .. \d\bq_{2l+1} \; \delta_D(\bk - \bq_1 - .. - \bq_{2l+1})  \nonumber \\ & & \times \; F_{2l+1}^{(s)}(\bq_1, .. , \bq_{2l+1}) \; \lag \dL(\bk') \dL(\bq_1) .. \dL(\bq_{2l+1}) \rag .
\label{P1lk1}
\eeqa
Using eq.(\ref{PL1}) and Wick's theorem we obtain:
\beqa
P_{1,2l+1}(k) & = & (2l+1)!! \; \int \d\bq_1 .. \d\bq_l \; P_L(k) P_L(q_1) .. P_L(q_l) \nonumber \\ & & \times \; F_{2l+1}^{(s)}(\bk,\bq_1,-\bq_1,..,\bq_l,-\bq_l)  .
\label{P1lk2}
\eeqa
As we recalled in Sect.\ref{Equations of motion} the kernels $F_{l}^{(s)}$ are proportional to $k^2$ for $l\geq 2$ (e.g., \cite{Goroff1}) because of momentum conservation. Therefore, introducing a high-wavenumber cutoff $\Lambda$ (i.e. $P_L(k) =0$ for $k>\Lambda$) we have:
\beq
P_{1,2l+1}(k) \sim P_L(k) \left( \frac{k}{\Lambda} \right)^2 \left[ \Lambda^3 P_L(\Lambda) \right]^l
\label{P1lk3}
\eeq
provided the integration over $\bq_1,..,\bq_l$ diverges at large wavenumbers. For power-law power-spectra this condition corresponds to:
\beq
(3+n) l - 2 \geq 0 .
\label{Lamb1}
\eeq
Therefore, all such diagrams with $l \geq 2/(3+n)$ yield a contribution $k^2 P_L(k)$ to the non-linear power-spectrum $P(k)$.

Here, we investigate the large-scale limit, which is different from the early-time limit (if the power-spectrum is not a pure power-law). This is the limit of interest for practical purposes, where one tries to describe the density field on large scales which are still linear while the small scales have already entered the non-linear regime. Thus, we have shown above that all diagrams with only one large branch of size $R$ and $2l+1$ factors $\dL$ at point $\bx$ or $\bx'$ (and one factor $\dL$ on the other side) in Fig.\ref{figexp1} yield a large-scale tail of order $k^2 P_L(k)$ for $P(k)$, if $l \geq 2/(3+n)$. This corresponds to a term $R^{-2} \sigma^2(R)$ for the non-linear two-point correlation $\xi(R)$ and variance $\xib(R)$. Here we introduced the rms linear density fluctuation $\sigma$ in a cell of radius $R$, volume $V$:
\beq
\sigma^2(R) \equiv \lag \delta_{L,R}^2 \rag = \int_V \frac{\d\bx_1}{V} \frac{\d\bx_2}{V} \; \DL(\bx_1,\bx_2) ,
\label{sig1}
\eeq
where $\delta_{L,R}$ is the linear density contrast within the spherical cell of radius $R$. The non-linear variance $\xib(R)$ is simply given by:
\beq
\xib(R) \equiv \lag \dR^2 \rag
\label{xib1}
\eeq
where $\dR$ is the non-linear density contrast within the spherical cell. Therefore, we see that the large-scale non-linear two-point correlation $\xi(R)$ cannot be expanded as a series over powers of the linear variance $\sigma^{2q}$, with $q\geq 1$. Indeed, beyond some order $q$ there appears a term of order $R^{-2} \sigma^2(R)$ arising from the diagrams described above. Note that diagrams with one branch of size $R$ and with $l$ (resp. $l'$) factors $\dL$ at the point $\bx$ (resp. $\bx'$) with both $l>1$ and $l'>1$ yield a contribution smaller than $R^{-2} \sigma^2(R)$ by additional powers of $\sigma^2$ or $R^{-2}$ (or an entire power of $\sigma^2$). On the other hand, the contribution from diagrams with $l$ branches of size $R$ (e.g., $l=2$ for the diagram (3) in Fig.\ref{figexp1}) exhibits at least a factor $\sigma^{2l}$. Therefore, the first ``non-regular'' contribution to the non-linear two-point correlation (i.e. not of the form $\sigma^{2q}$) is the term $R^{-2} \sigma^2(R)$ described above.

The order $q_+$ of the last ``regular term'' is given by the constraint $q_+(n+3) < 2+(n+3)$ which yields:
\beq
q_+ = 1 + \left[ \frac{2}{n+3} \right] ,
\label{qp1}
\eeq
where $[..]$ is the entire part. Therefore, we see that for all power-spectra the expansion of the non-linear correlation $\xi(R)$ as a series over powers of $\sigma^2$ breaks down beyond a finite order $q_+$. This order $q_+$ is larger for smaller $n$ which corresponds to initial conditions with less small-scale power. In particular, for $n \geq -1$ the only regular term in the expansion of $\xi$ (or $P(k)$) is the leading order-term $\sigma^2$ since $q_+=1$. This means that the so-called ``one-loop term'' shows a small-scale divergence. This agrees with the results of the previous studies (e.g., \cite{Lok1}, \cite{Scoc1}). For $n<-1$ the break up of the expansion over powers of $\sigma^2$ only occurs at a higher order. Thus, in this respect the index $n=-1$ does not correspond to a significant change of the physics. Here, we note that two different cases may occur with respect to the new contribution $R^{-2} \sigma^2(R)$ which arises from an infinite series of diagrams. 

Firstly, the high-wavenumber cutoff $\Lambda$ (in case it is set by the initial shape of the linear power-spectrum $P_L(k)$) may still be within the linear regime. Then, the perturbative expansion makes sense and one may still be able to write a perturbative expression for the large-scale non-linear two-point correlation $\xi(R)$. However, in addition to terms of the form $\sigma^{2q}$ one must take into account the new contributions like $R^{-2} \sigma^2(R)$ which arise from particular classes of diagrams. This requires some partial resummations over infinite series of diagrams which may still lead to finite results (a definite statement requires an explicit calculation). If we also expand over powers of $\Lambda^3 P_L(\Lambda) \sim \sigma^2(1/\Lambda) \ll 1$ we obtain a finite number of diagrams, up to the required orders $\sigma(R)^{2q_1}$ and $\sigma(1/\Lambda)^{2q_2}$. Note that in any case this implies that the physics at scale $R$ also depends on the shape of the linear power-spectrum at the much smaller scale $1/\Lambda$. In fact, this is not totally surprising. Indeed, although we could expect the large-scale behaviour to be ``largely'' independent of the small-scale physics, this decoupling is not exact. That is, by choosing different cutoffs $\Lambda$ we clearly obtain different density fields and this must also show up, to some degree, in the exact properties of the density field at large scale $R$, even if $R \gg 1/\Lambda$. Thus, our analysis shows that in the limit $R \ll 1/\Lambda \ll R_0$, where $R_0$ is the comoving scale associated with the transition to the non-linear regime (defined for instance by $\sigma(R_0)=1$), the dependence on $\Lambda$ does not show up in the leading-order term of $\xib$ (which is merely $\sigma^2$) but it can be traced in higher-order corrections. This clearly agrees with intuitive expectations.

Secondly, the cutoff $\Lambda$ may already be within the non-linear regime. This is actually the case of cosmological interest where the smaller scales have usually turned non-linear at the time of interest. Then, it is clear that the perturbative approach fails and rigorous results require a non-perturbative method. Thus, the perturbative contributions which are dominated by the cutoff $\Lambda$, as in eq.(\ref{P1lk3}), cannot yield correct results since large wavenumbers in the range $1/R_0 < k < \Lambda$ cannot be described by perturbative means. Indeed, on these small scales one must take into account shell-crossing, which is out of reach of any approach which is based on the hydrodynamical description. One might still be able to estimate the order of magnitude of the terms like eq.(\ref{P1lk3}) by setting $\Lambda=1/R_0$, in the sense that the non-linear dynamics itself may generate the required cutoff to make the diagrams finite. However, it is clear that this heuristic procedure requires some non-perturbative calculations in order to get any convincing support. Besides, the divergence of subleading terms casts some doubts on the validity of the perturbative results obtained at lower orders. We shall come back to these points in Sect.\ref{Steepest-descent method} and  Sect.\ref{Comparison with simple dynamics} below.

Finally, we note from eq.(\ref{P1lk3}) that the degree of divergence of the terms encountered in the perturbative expansion can be made arbitrarily large since it increases at higher orders, as $\Lambda^{(3+n)l-2}$. Therefore, this is a non-renormalizable theory and standard renormalization procedures cannot be applied (e.g., \cite {Zinn1}).

\section{Steepest-descent method}
\label{Steepest-descent method}

In a previous article (\cite{paper2}) we developed a steepest-descent method to study the properties of the density field in the quasi-linear regime. Here we describe how the analysis performed in the previous section from the standard perturbative approach translates within this new framework. We study the statistics of the density contrast $\dR$ within a spherical cell of comoving radius $R$, volume $V$:
\beq
\dR = \int_V \frac{\d^3x}{V} \; \delta(\bx) .
\label{dR}
\eeq
To this order, we introduce the generating function $\psib(y)$ defined by the relation:
\beq
\psib(y) \equiv \lag e^{-y \dR / \sigma^2(R) } \rag \equiv \int_{-1}^{\infty} \d\dR \; e^{-y \dR / \sigma^2(R) } \; \cP(\dR) .
\label{psib1}
\eeq
The average $\lag .. \rag$ in eq.(\ref{psib1}) denotes the mean over the initial conditions, which are defined by the linear growing mode $\dL(\bx)$. The last equality defines the probability distribution function (pdf) $\cP(\dR)$ of the density contrast. Thus, $\psib(y)$ is also the ``rescaled'' Laplace transform of $\cP(\dR)$. It also yields the moments of the density contrast $\dR$ through the expansion:
\beq
\psib(y) = \sum_{q=0}^{\infty} \frac{(- y)^q}{q!} \; \frac{\lag \dR^q \rag}{\sigma^{2q}} .
\label{psib2}
\eeq
It is convenient to introduce also the generating function $\varphi(y)$ of the cumulants defined by:
\beq
\psib(y) \equiv e^{-\varphi(y)/\sigma^2} ,
\label{phi1}
\eeq
which exhibits the expansion:
\beq
\varphi(y) = - \sum_{q=2}^{\infty} \frac{(- y)^q}{q!} \; \frac{\lag \dR^q \rag_c}{\sigma^{2(q-1)}} ,
\label{phi2}
\eeq
since we have $\lag \dR \rag_c=0$. Thus, the term of order $y^2$ of the expansion over $y$ of the generating functions $\psib(y)$ or $\varphi(y)$ yields the non-linear variance $\xib=\lag \dR^2\rag_c$. 

Then, as shown in \cite{paper2} the generating function $\psib(y)$ can be expressed as a path-integral through the relation:
\beq
\psib(y) = \left( \Det\DL^{-1} \right)^{1/2} \int [\d\dL(\bx)] \; e^{- S[\dL] /\sigma^2(R)}
\label{psi3}
\eeq
where we introduced the action $S[\dL]$:
\beq
S[\dL] \equiv y \; \dR[\dL] + \frac{\sigma^2(R)}{2} \; \dL . \DL^{-1} . \dL
\label{S0}
\eeq
Here $\dR[\dL]$ is the non-linear density contrast $\dR$ within the spherical cell $V$ obtained for a linear density field $\dL(\bx)$. The relation (\ref{psi3}) is merely the explicit expression of the Gaussian average over the linear growing mode $\dL(\bx)$ which enters the definition of the generating function $\psib(y)$ in eq.(\ref{psib1}).

The action $S[\dL]$ is independent of the normalization of the power-spectrum $P_L(k)$ since $\DL \propto \sigma^2$. Then, it is clear that the path-integral in eq.(\ref{psi3}) is dominated by the minimum of the action $S$ in the limit $\sigma \rightarrow 0$ for a fixed $y$. Indeed, the contributions from other states $\dL$ are exponentially damped relative to this point. Moreover, the steepest-descent approximation becomes exact in this limit. Here we must point out that this actually corresponds to the ``early-time'' limit and not to the ``large-scale'' limit. Indeed, the shape of the linear power-spectrum $P_L(k)$ (or of the kernel $\DL$) is held fixed while its amplitude goes to zero.

Next, let us note $\dy$ the saddle-point of the action $S$, for a given $y$, and $S_y = S[\dy]$ the value of the action at this point. Expanding the action around the point $\dy$ we write:
\beq
S[\dy + \delta_L'] = S_y + \frac{1}{2} W_2 . \dL'^2 + \frac{1}{3!} W_3 . \dL'^3 + .. ,
\label{S1}
\eeq
where we introduced the symmetric kernels $W_q$ defined by the functional derivatives at the point $\dy$:
\beq
W_q(\bx_1,..,\bx_q) \equiv \left. \frac{\delta^q S}{\delta(\delta_L(\bx_1)) .. \delta(\delta_L(\bx_q))} \right|_{\dy} ,
\label{Wq1}
\eeq
and we used the short-hand notation:
\beq
W_q . \dL'^q \equiv \int \d\bx_1 .. \d\bx_q \; W_q(\bx_1,..,\bx_q) \; \dL'(\bx_1) .. \dL'(\bx_q) .
\label{Wq2}
\eeq
As noticed in \cite{paper2}, for small $y$ the kernel $W_2$ is definite positive since $W_2(y=0)= \sigma^2 \DL^{-1}$, see eq.(\ref{S0}). Then, we can use the steepest-descent method in the limit $\sigma \rightarrow 0$. The quadratic term $(W_2 . \dL'^2)/\sigma^2$ in the exponential in eq.(\ref{psi3}) provides a cutoff of order $\sigma$ for $\dL'$ so that the next term $(W_3 . \dL'^3)/\sigma^2$ is of order $\sigma \rightarrow 0$. This confirms that at leading order the generating function $\psib(y)$ is given by the Gaussian integration around the saddle-point $\dy$. This yields:
\beq
\psib(y) = \left( \Det\DL^{-1} \right)^{1/2} \; \left( \Det M_y \right)^{-1/2} \; e^{- S_y /\sigma^2(R)} + ...
\label{psi4}
\eeq
where the matrix $M_y$ is the Hessian of the exponent at the point $\dy$:
\beqa
\lefteqn{ M_y(\bx_1,\bx_2) \equiv \left. \frac{\delta^2 (S/\sigma^2)}{\delta(\dL(\bx_1)) \delta(\dL(\bx_2))} \right|_{\dy} } \nonumber \\ & & = \left. \frac{y}{\sigma^2(R)} \; \frac{\delta^2 (\dR)}{\delta(\dL(\bx_1)) \delta(\dL(\bx_2))} \right|_{\dy} +  \DL^{-1}(\bx_1,\bx_2) . 
\label{M1}
\eeqa
Higher-order corrections to $\psib(y)$ (i.e. higher powers of $\sigma^2$) are obtained by expanding the terms of order $q \geq 3$ of the action $S$ in the exponential in eq.(\ref{psi3}). In fact, the expression (\ref{psi4}) already gives the next-to-leading order term of the cumulants $\lag \dR^q \rag_c$. Indeed, from eq.(\ref{phi1}) we have:
\beq
\varphi(y) = - \sigma^2 \ln \psib = S_y + \frac{\sigma^2}{2} \ln \Det (\DL.M_y) + ..
\label{phi3}
\eeq
This provides the generating function $\varphi(y)$ as an expansion over $\sigma^2$. As explained above, the higher-order terms in eq.(\ref{phi3}) come from the expansion of the exponential of the action beyond its quadratic term in eq.(\ref{psi3}). Here we must point out that although the steepest-descent method yields an expansion over $\sigma^2$ it can actually go beyond the reach of the standard perturbative approach described in Sect.\ref{Standard perturbative expansion}. For instance, as shown in \cite{paper2}, for power-spectra with $n<0$ this method yields the correct slow decline at large densities of the pdf $\cP(\dR)$ (shallower than a mere exponential) which cannot be rigorously derived from the usual perturbative approach. This corresponds to the case where the radius of convergence of the series expansion (\ref{phi2}) is zero. The reason why it is possible to go beyond the standard perturbative results is that the saddle-point $\dy$ is obtained in a non-perturbative way (this is made possible by the fact that the action $S[\dL]$ is spherically symmetric and we know the explicit solution of the spherical dynamics before shell-crossing, see \cite{paper2}). Besides, as shown in \cite{paper4} the steepest-descent method can also be applied to the non-linear regime where it provides the tails of the pdf $\cP(\dR)$.

For practical purposes, it is convenient to write eq.(\ref{phi3}) as:
\beq
\varphi(y) = S_y + \frac{\sigma^2}{2} \Tr \ln (1+A_y) + ..
\label{A1}
\eeq
where we defined the matrix $A_y$ by:
\beq
A_y(\bx_1,\bx_2) \equiv \frac{y}{\sigma^2} \int \d\bx \; \DL(\bx_1,\bx) \left. \frac{\delta^2(\dR)}{\delta(\dL(\bx)) \delta(\dL(\bx_2))} \right|_{\dy} 
\label{A2}
\eeq
The value $S_y$ of the action at the saddle-point $\dy$ was derived in \cite{paper2}. It yields the leading-order contribution to the generating function $\varphi(y)$ and to the cumulants $\lag \dR^q \rag_c$, through eq.(\ref{phi2}). In this article we are interested in the subleading corrections. In particular, the next-to-leading contribution is given by the term of order $\sigma^2$ written in eq.(\ref{A1}). In order to derive the correction associated with a cumulant of order $q$ we simply need to expand the logarithm $\ln (1+A_y)$ as a power series over $y$, up to order $y^q$. Here, we shall only consider the cumulant of order 2 (that is the power-spectrum or the two-point correlation) so that we can write eq.(\ref{A1}) as:
\beq
y \lag \dR \rag_c - \frac{y^2}{2} \frac{\lag \dR^2 \rag_c}{\sigma^2} + .. = - \frac{y^2}{2} + \frac{\sigma^2}{2} \left( \Tr A_y - \frac{1}{2} \Tr A_y^2 \right) + ..
\label{A3}
\eeq
where we used $S_y=-y^2/2 + {\cal O}(y^3)$ (see \cite{paper2}) and the fact that $A_y$ is of order $y$ (see eq.(\ref{A2})). Note that we must check that we recover $\lag \dR \rag_c=0$. We shall also recover the results of Sect.\ref{Standard perturbative expansion}.

Now, we need the expression of the matrix $A_y$ up to order $y^2$. Therefore, we must obtain the second-derivative which appears in eq.(\ref{A2}) up to order $y$. Moreover, as seen in \cite{paper2} the saddle-point $\dy$ is of order $y$ in the limit $y \rightarrow 0$, see also eq.(\ref{dy1}) below, so that we can use the expansion (\ref{Flx1}) to write up to order $y$:
\beqa
\lefteqn{ \delta(\bx)[\dy + \dL'] = \int \d\bx_1 \d\bx_2 \; F_2^{(s)}(\bx;\bx_1,\bx_2) \; \dL'(\bx_1) \dL'(\bx_2) } \nonumber \\ & & + 3 \int \d\bx_1 .. \d\bx_3 F_3^{(s)}(\bx;\bx_1,..,\bx_3) \dL'(\bx_1) \dL'(\bx_2) \dy(\bx_3)
\label{dRy1}
\eeqa
where we only kept the terms of order $\dL'^2$. Hence we have up to order $y^2$:
\beqa
\lefteqn{ A_y(\bx_1,\bx_2) = \frac{2 y}{\sigma^2} \int_V \frac{\d\bx}{V} \int \d\bx' \; \DL(\bx_1,\bx') F_2^{(s)}(\bx;\bx',\bx_2) } \nonumber \\ & & + \frac{6 y}{\sigma^2} \int_V \frac{\d\bx}{V} \int \d\bx' \d\bx'' \; \DL(\bx_1,\bx') F_3^{(s)}(\bx;\bx',\bx'',\bx_2) \dy(\bx'') \nonumber \\ & & + ..
\label{A4}
\eeqa

From eq.(\ref{A3}) we obtain for the cumulant of order 1, up to order $\sigma^2$:
\beq
\lag \dR \rag_c = \int_V \frac{\d\bx}{V} \int \d\bx' \d\bx'' \DL(\bx'',\bx') F_2^{(s)}(\bx;\bx',\bx'') .
\label{dRc1}
\eeq
Using the standard result (e.g., \cite{Peebles1}):
\beq
F_2^{(s)}(\bk_1,\bk_2) = \frac{5}{7} + \frac{1}{2} \frac{\bk_1.\bk_2}{k_1^2} + \frac{1}{2} \frac{\bk_1.\bk_2}{k_2^2} + \frac{2}{7} \left( \frac{\bk_1.\bk_2}{k_1 k_2} \right)^2
\label{F2s1}
\eeq
it is easy to show that we recover $\lag \dR \rag_c =0$. Then, from \cite{paper2} we can write the saddle-point $\dy(\bx)$ up to order $y$ as:
\beq
\dy(\bx) = - \frac{y}{\sigma^2} \int_V \frac{\d\bx'}{V} \; \DL(\bx,\bx') + \cO(y^2) .
\label{dy1}
\eeq
This yields for the cumulant of order 2, up to order $\sigma^4$:
\beqa
\lefteqn{ \lag \dR^2 \rag_c = \sigma^2 } \nonumber \\ & & + \; 6 \int_V \frac{\d\bx \d\bx'}{V^2} \int \d\bx_1 .. \d\bx_3 \; F_3^{(s)}(\bx;\bx_1,\bx_2,\bx_3) \nonumber \\  & &  \hspace{1cm} \times \; \DL(\bx_1,\bx_2) \DL(\bx_3,\bx') \nonumber \\ & & + \; 2 \int_V \frac{\d\bx \d\bx'}{V^2} \int \d\bx_1 .. \d\bx_4 \; F_2^{(s)}(\bx;\bx_1,\bx_2) F_2^{(s)}(\bx';\bx_3,\bx_4) \nonumber \\  & & \hspace{1cm} \times \; \DL(\bx_1,\bx_3) \DL(\bx_2,\bx_4) .
\label{dRc2}
\eeqa
Thus, we recover the result we would obtain from the standard expansion displayed in Fig.\ref{figexp1}. This is not surprising since we used the same expansion (\ref{Flx1}) to write the fluctuations of the functional $\dR[\dL]$ in eq.(\ref{A4}).

Of course, as in Sect.\ref{Small scale divergences}, in the case of power-law power-spectra we must add a high-wavenumber cutoff $\Lambda$ in order to obtain finite results. Note that within this steepest-descent method we consider the ``early-time'' limit where the amplitude of the power-spectrum goes to zero, and not the ``large-scale'' limit. Therefore, $\sigma^2(1/\Lambda)$ is of the same order as $\sigma^2(R)$ (they are both proportional to the amplitude of the linear power-spectrum, whose shape is held fixed) and the expansion obtained from eq.(\ref{psi3}) is over powers of both $\sigma^2(1/\Lambda)$ and $\sigma^2(R)$, which are treated on the same footing. In the limit $\Lambda \rightarrow \infty$ we again encounter divergent quantities, as in Sect.\ref{Small scale divergences}.

In order to go beyond these divergences we should use a non-perturbative expression for the functional $\dR[\dL]$. In particular, perturbative expressions like eq.(\ref{dRy1}) for the fluctuations of the non-linear density contrast $\dR$ around the saddle-point $\dy$ are not sufficient. However, we note that the saddle-point $\dy$ and the value $S_y$ of the action at this point are obtained in a non-perturbative manner. Indeed, as shown in \cite{paper2} this calculation does not rely on a perturbative expansion for the density field $\delta(\bx)$. Therefore, the quantity $S_y$ is the exact minimum of the action $S[\dL]$, even if we use the exact non-linear expression for the functional $\dR[\dL]$. Hence the leading-order term $S_y$ for the generating function $\varphi(y)$ is correct and it is not affected by the divergence of subleading terms beyond some finite order. From the expansion (\ref{phi2}) this also means that the value obtained for the cumulants at leading-order (``tree-level'') is exact. 

This agrees with the fact that numerical simulations are seen to match the predictions of leading-order perturbation theory which is consistent with eq.(\ref{A1}) for the cumulants $\lag \dR^q \rag_c$, see \cite{Ber2} and \cite{paper2}. This result could also be expected within the framework of the usual perturbation theory. Indeed, as shown in Bernardeau (1992,1994), at leading-order in the limit $\sigma \rightarrow 0$ the cumulants only involve ``tree diagrams'' with no ``small internal branches'' which were at the origin of the small-scale divergences which appeared for instance in eq.(\ref{P1lk3}).

As discussed below in Sect.\ref{The one-dimensional Burgers equation}, we can actually expect the results of perturbation theory to be valid up to the order $q_+$ obtained in eq.(\ref{qp1}), that is until they are finite.

\section{Comparison with two simple dynamics}
\label{Comparison with simple dynamics}

We have seen in the previous sections that for power-law power-spectra $P_L(k) \propto k^n$ one always encounters some divergent quantities beyond some finite order in a perturbative expansion over $\sigma^2$. However, in the quasi-linear regime the results from numerical simulations agree very well with the leading-order terms obtained by such analytical means (e.g., \cite{Ber2}). Of course, numerical simulations always have a small-scale cutoff due to finite resolution effects but it has been checked that the behaviour they obtain converges and becomes independent of this cutoff (i.e., the limit $\Lambda \rightarrow \infty$ appears to be well-defined). This suggests that the divergence of these subleading terms does not spoil the predictions obtained from the calculations of the lower-order terms. 

One may then argue that the non-linear dynamics itself could provide a small-scale cutoff $R_0$. Indeed, at scales smaller than $R_0$ which marks the transition to the non-linear regime ($\sigma(R_0)=1$) perturbative approaches break down so that the integrals obtained in the previous sections do not make sense at large wavenumbers $k > 1/R_0$. Therefore, one should cut the integrations at $\Lambda = 1/R_0$ and the contributions of higher wavenumbers should be derived from a fully non-perturbative method. Then, this could provide finite results (see also \cite{Jain2}, \cite{Scoc1}, for discussions on this point). Unfortunately, a more detailed statement requires explicit non-perturbative calculations which remain to be done. In order to get some insight into this problem, it could be useful to study a somewhat simpler dynamics as described below.

\subsection{The Zel'dovich approximation}
\label{The Zel'dovich approximation}

For instance, we may try to investigate these small-scale divergences within the framework of the Zel'dovich approximation (\cite{Zel1}). Unfortunately, we shall see below that these divergences actually disappear in this simpler dynamics, if it is to make any sense. Indeed, as shown in \cite{Gri1}, within the framework of the Zel'dovich approximation (ZA) the symmetric kernels $F_l^{(s)}$ which enter the expansion (\ref{exp1}) through eq.(\ref{Fl1}) are given by:
\beq
{\rm (ZA)}: \;\; F_l^{(s)}(\bq_1, .. , \bq_l) = \frac{1}{l!} \; \frac{\bk.\bq_1}{q_1^2} .. \frac{\bk.\bq_l}{q_l^2} .
\label{FsZA1}
\eeq
Therefore, we have $F_{2l+1}^{(s)}(\bk,\bq_1,-\bq_1, .. ,\bq_l,-\bq_l)  \propto k^{2l}$ and eq.(\ref{P1lk3}) becomes:
\beqa
{\rm (ZA)}: \;\; P_{1,2l+1}(k) & \sim & P_L(k) \left( \frac{k}{\Lambda} \right)^{2l} \left[ \Lambda^3 P_L(\Lambda) \right]^l \nonumber \\ & \propto & \Lambda^{(n+1)l} 
\label{P1lZA1}
\eeqa
provided the integral is dominated by the cutoff $\Lambda$. We see that this constraint requires $n>-1$. However, for such power-spectra we cannot use the Zel'dovich approximation because the linear peculiar velocity field $\bv_L(\bx)$ is not well-defined. That is, high wavenumbers $k$ provide a divergent contribution to the unsmoothed velocity (but of course the mean velocity over some finite spherical cell $V$ is finite). Therefore, we cannot study this dynamics without imposing some high-wavenumber cutoff $\Lambda$. Note that one sometimes uses the non-linear scale $R_0$ as a small-scale cutoff, which is refered to as the ``truncated Zel'dovich approximation'' (e.g., \cite{Col1}). On the other hand, for $n < -1$ there are no small-scale divergences at all: all diagrams are finite. This allows one to write a perturbative expansion for the non-linear two-point correlation $\xi(R)$ where all terms are finite (\cite{Bha1}). Therefore, the Zel'dovich and the exact gravitational dynamics exhibit rather different behaviours in this respect. Thus, we cannot use the simple Zel'dovich approximation to investigate the small-scale divergences. However, we can use the adhesion model which still makes sense for $n>-1$, as described below.

\subsection{The one-dimensional Burgers equation}
\label{The one-dimensional Burgers equation}

The adhesion model allows one to mimic the gravitational sticking of particles after shell-crossing which is not described by the Zel'dovich approximation, see \cite{Gur1}. This provides a reasonable description of the large-scale skeleton of gravitational structures (e.g., \cite{Verg1}). To do so, one simply adds a vanishing viscosity to the Zel'dovich equations of motion. Then, as shown in \cite{Gur1} (see also \cite{Verg1}) the evolution of the velocity field (after suitable rescaling) is described by the standard Burgers equation. The infinitesimal viscosity prevents shell-crossing as particles stick together after collision. Then, contrary to the simple Zel'dovich approximation discussed in Sect.\ref{The Zel'dovich approximation} we can study initial conditions with $n>-1$. Indeed, the non-linear dynamics itself makes the velocity field finite at any time $t>0$. Here we are only interested in the adhesion model for illustrative purposes hence we consider the standard one-dimensional Burgers equation:
\beq
\frac{\pl v}{\pl t} + v \frac{\pl v}{\pl x} = \nu \frac{\pl^2 v}{\pl x^2} ,
\label{Bur1}
\eeq
in the inviscid limit $\nu \rightarrow 0$. Moreover, we investigate the case of a white-noise power-spectrum $P_{v_0}(k)$ for the initial Gaussian velocity field $v_0(x)$ at $t=0$ (i.e. $P_{v_0}(k)$ is flat):
\beqa
\lag v_0(k_1) v_0(k_2) \rag & \equiv & P_{v_0}(k_1) \; \delta_D(k_1+k_2) \nonumber \\ & = & x_0 \left( \frac{x_0}{t_0} \right)^2 \delta_D(k_1+k_2) .
\eeqa
Here the length $x_0$ and the time $t_0$ are parameters which merely define the normalization of the power-spectrum $P_{v_0}(k)$. Then, the two-point correlation of the initial velocity field is:
\beqa
\Delta_{v_0}(x_1,x_2) & \equiv & \lag v_0(x_1) v_0(x_2) \rag \nonumber \\ & = & 2\pi \; x_0 \left( \frac{x_0}{t_0} \right)^2\delta_D(x_1-x_2) .
\label{Dv0}
\eeqa
The kernel $\Delta_{v_0}$ fully defines the statistics of the Gaussian random field $v_0(x)$. Then, the velocity field $v(x,t)$ at time $t>0$ is obtained by solving Burgers eq.(\ref{Bur1}) with the initial condition $v_0(x)$. This classical problem is studied in details in \cite{Burg1} (see also \cite{Gur1} and \cite{Verg1}). In particular, the exact solution $v(x,t)$ can be obtained through the standard Hopf-Cole transformation.

Starting at $t=0$ with a uniform density distribution it is easy to see that at any regular point $x$ the density contrast $\delta(x,t)$ is given by:
\beq
\delta(x) = - t \frac{\pl v}{\pl x} .
\label{delv}
\eeq
Therefore, in order to obtain the statistics of the density field we merely need the properties of the velocity field. First, we can investigate a perturbative approach in order to derive $v(x,t)$, in a fashion similar to the procedure detailed in Sect.\ref{Standard perturbative expansion} for the gravitational dynamics. From eq.(\ref{Bur1}) we obtain at linear order in the inviscid limit ($\nu \rightarrow 0$) the simple relation $\pl v/\pl t=0$. Thus, the linear solutions $v_L(x,t)$ and $\dL(x,t)$ are:
\beq
v_L(x,t) = v_0(x) , \;\; \dL(x,t) = - t \frac{\pl v_0}{\pl x} .
\label{vdL1}
\eeq
Then, as in eq.(\ref{exp1}) we can look for a perturbative solution $v(x,t)$ and $\delta(x,t)$, written as an expansion over the initial velocity field $v_0(x)$ (i.e. over the linear growing mode $\dL(x,t)$). Shell-crossing does not appear in the perturbative approach (e.g., \cite{Bha1}, \cite{paper1}) so that in the limit $\nu \rightarrow 0$ we actually recover the Zel'dovich dynamics (i.e. the r.h.s. in eq.(\ref{Bur1}) disappears as $\nu \rightarrow 0$ within perturbation theory). Then eq.(\ref{Bur1}) reads $\d v/\d t=0$ and the distribution function $f(x,v;t)$ is simply:
\beq
f(x,v;t) = \rho_0 \; \delta_D(v-v_0(s)) \;\; \mbox{with} \;\; x(s,t) = s + t v ,
\label{fZel1}
\eeq
where $\rho_0$ is the initial uniform density. Next, expanding the Dirac in eq.(\ref{fZel1}) and using $\rho(x,t) = \int f(x,v;t) \d v$ we obtain as in \cite{Gri1}:
\beq
\delta(x,t) = \sum_{l=1}^{\infty} \frac{(-1)^l}{l!} \; t^l \; \frac{\pl^l}{\pl x^l} \left[ v_0(x)^l \right] .
\label{delexp1}
\eeq
In Fourier space this yields:
\beqa
\delta(k) & = & \sum_{l=1}^{\infty} \frac{1}{l!} \int \d q_1 .. \d q_l \; \delta_D(k - q_1 - .. - q_l) \nonumber \\ & & \times \; \frac{k^l}{q_1 .. q_l} \; \dL(q_1) .. \dL(q_l) .
\label{delexp2}
\eeqa
Hence we recover the one-dimensional form of eq.(\ref{FsZA1}) for the kernels $F_l^{(s)}(q_1, .. ,q_l)$. From eq.(\ref{vdL1}) the power-spectrum $P_L(k,t)$ of the linear density field $\dL(x,t)$ is simply:
\beq
P_L(k,t) = t^2 k^2 P_{v_0}(k) \propto k^2
\label{PLB1}
\eeq
so that eq.(\ref{P1lZA1}) becomes:
\beq
P_{1,2l+1}(k) \sim P_L(k) \left( \frac{k}{\Lambda} \right)^{2l} \left[ \Lambda P_L(\Lambda) \right]^l \propto \Lambda^l .
\label{PLB2}
\eeq
Thus, we see that we again encounter divergent quantities in perturbation theory, as in Sect.\ref{Small scale divergences} for the gravitational dynamics. Note that these divergences already appear at the next-to-leading order, for $l=1$ in eq.(\ref{PLB2}).

In fact, for a white-noise initial velocity the density contrast within a top-hat of radius $R$ is not well-defined, even at linear order. Indeed, the initial velocity field $v_0(x)$ is not a smooth function but a distribution. Thus, we study the mean density contrast $\dRG$ over a Gaussian window of size $R$:
\beq
\dRG \equiv \int_{-\infty}^{\infty} \frac{\d x}{\sqrt{2\pi}R} \; e^{-x^2/(2R^2)} \; \delta(x) .
\label{dRG1}
\eeq
The Gaussian cutoff $\sim e^{-(kR)^2/2}$ at high wavenumbers ensures that $\lag (\dRG)^2 \rag$ is well-defined at linear order. By contrast, a top-hat window exhibits an extended power-law tail at large $k$. Thus, in order to really probe the physics at scale $R$ it is better to use the Gaussian window. Using eq.(\ref{delv}) we can write:
\beq
\dRG = \frac{- t}{\sqrt{2\pi}R^3} \int \d x \; e^{-x^2/(2R^2)} \; x \; v(x)
\label{dRG2}
\eeq
and:
\beq
\lag (\dRG)^2 \rag = \frac{t^2}{4\sqrt{\pi}R^3} \int \d x \; e^{-x^2/(4R^2)} \left( 1 - \frac{x^2}{2R^2} \right) \Delta_v(x) .
\label{dRG3}
\eeq
Here we introduced the two-point velocity correlation $\Delta_v(x) \equiv \lag v(r)v(r+x)\rag$. At linear order we have $v=v_L=v_0$, hence $\Delta_v(x)=\Delta_{v0}(x)$ and using eq.(\ref{Dv0}) we obtain:
\beq
\sigma_G^2(R) \equiv \lag (\delta_{L,R}^G)^2 \rag = \frac{\sqrt{\pi}}{2} \; \left( \frac{x_0}{R} \right)^3 \; \left( \frac{t}{t_0} \right)^2 .
\label{siG1}
\eeq
Note that small scales turn non-linear first, following a hierarchical process as for the usual cosmological scenario of structure formation. Then, a perturbative approach would give the non-linear variance $\lag (\dRG)^2 \rag$ as an expansion over powers of $\sigma_G^2$. However, as noticed from eq.(\ref{PLB2}) the perturbative method already fails at the next-to-leading level.

The advantage of the Burgers equation (\ref{Bur1}) is that it can be solved explicitly through the Hopf-Cole transformation. This allows one to derive numerous exact results in a non-perturbative way (e.g., \cite{Burg1} and references in \cite{Verg1}). In particular, we can obtain the variance $\lag (\dRG)^2 \rag$ as an expansion over $1/R$ in the large-scale limit $R \rightarrow \infty$ by expanding the exponential in eq.(\ref{dRG3}). At leading order we get:
\beq
\lag (\dRG)^2 \rag^{(1)} = \frac{t^2}{4\sqrt{\pi}R^3} \int_{-\infty}^{\infty} \d x \; \Delta_v(x) .
\label{dRGl1}
\eeq
Next, we note that the integral $J \equiv \int \d x \Delta_v(x)$ is actually independent of time (i.e. it is an invariant, see \cite{Burg1}), as can be seen from eq.(\ref{Bur1}). Using eq.(\ref{Dv0}) we obtain:
\beq
J \equiv \int \d x \; \Delta_v(x) = \int \d x \; \Delta_{v0}(x) = 2\pi \; x_0 \left( \frac{x_0}{t_0} \right)^2 .
\label{J1}
\eeq
Substituting this result into eq.(\ref{dRGl1}) yields:
\beq
\lag (\dRG)^2 \rag^{(1)} = \frac{\sqrt{\pi}}{2} \; \left( \frac{x_0}{R} \right)^3 \; \left( \frac{t}{t_0} \right)^2 = \sigma_G^2(R) .
\label{dRGl2}
\eeq
Thus, at leading order in the large-scale limit $R \rightarrow \infty$ we recover the result (\ref{siG1}) of perturbation theory. In a similar fashion, the next-to-leading order term is:
\beq
\lag (\dRG)^2 \rag^{(2)} = \frac{-3 t^2}{16\sqrt{\pi}R^5} \int_{-\infty}^{\infty} \d x \; x^2 \; \Delta_v(x) \propto R^{-5} .
\label{dRGl3}
\eeq
Indeed, the two-point correlation $\Delta_v(x)$ vanishes as $\sim e^{-|x|^3}$ at large scales $|x| \rightarrow \infty$, see \cite{Fra1}, so that the integral converges. Thus, we can see that the term $\lag (\dRG)^2 \rag^{(2)}$ decreases more slowly than $\sigma_G^4$ at large scales. This explains why the perturbative approach must break down at the next-to-leading order. In fact, we see that the expansion of $\lag (\dRG)^2 \rag$ is of the form:
\beq
\lag (\dRG)^2 \rag = \sum_{l=1}^{\infty} \lag (\dRG)^2 \rag^{(l)} , \; \mbox{with} \; \lag (\dRG)^2 \rag^{(l)} \propto R^{-3-2(l-1)} .
\label{dRGl4}
\eeq
This expansion is not over powers of $\sigma_G^2$. In fact, beyond the order $q_+=1$ where perturbation theory is valid there are no more traces of the naive perturbative expansion since no powers of $\sigma_G^2$ appear. On the other hand, we note that the first ``non-regular'' term (i.e. which is not a power of $\sigma_G^2$) of the exact expansion (\ref{dRGl4}), obtained in eq.(\ref{dRGl3}), is actually of the form (\ref{PLB2}), with $l=1$. Thus, the non-linear dynamics has introduced a cutoff $\Lambda=1/R_0$, where $R_0$ is the scale which marks the transition to the non-linear regime. This makes the first divergent term (\ref{PLB2}) finite. However, this is not the whole story since the form of the exact large-scale expansion (\ref{dRGl4}) beyond this order cannot be infered from the naive perturbative theory (\ref{delexp1}) and (\ref{delexp2}), which would also give terms of the form $\sigma_G^{2l_1} R^{-2l_2}$ which do not appear in the exact result (\ref{dRGl4}).

We can reasonably expect the previous results obtained for the Burgers equation to apply to the case of the gravitational dynamics described in Sect.\ref{Small scale divergences}, although a definite statement would require an exact calculation. Indeed, the Burgers dynamics has been seen to provide a good description of gravitational clustering on large-scales (e.g., \cite{Gur1}, \cite{Verg1}). Moreover, the physics involved in the divergences of the perturbative expansions appears to be similar. Thus, in both cases the break down of the perturbative approach is related to its failure to take into account shell-crossing. Moreover, when particles cross each other caustics appear where the density is infinite and it is clear that perturbative methods cannot go beyond these points. Then, in the adhesion model particles stick together while for the exact gravitational dynamics particles actually cross each other and build ``virialized'' objects of finite thickness stabilized by a non-zero velocity dispersion. Nevertheless, in both cases the dynamics provides a cutoff at the non-linear scale $R_0$. For instance, while the typical displacement of particles diverges in perturbation theory, because of the contribution of high wavenumbers, it is clear that the actual contribution of small scales ($R < R_0$) is finite and of order $R_0$.

Thus, by analogy with eq.(\ref{dRGl2}) we can expect the perturbative results obtained through the standard expansions (\ref{exp1}) to be valid in the large-scale limit up to the order $q_+$ derived in eq.(\ref{qp1}), that is until they are finite. This agrees both with the analysis described in Sect.\ref{Steepest-descent method} which shows that the leading-order terms must be correct and with the calculations performed in \cite{Scoc1} which find that numerical simulations match the predictions of perturbation theory up to the next-to-leading order for $n<-1$, when they are finite. Moreover, by analogy with eq.(\ref{PLB2}) and eq.(\ref{dRGl3}), we can expect from eq.(\ref{P1lk3}) that the first ``non-regular'' term (i.e. not of the form $\sigma^{2q}$) in the expansion of $\xib(R)$ over powers of $1/R$ in the large-scale limit is of the form $R^{-2} \sigma^2(R)$. Finally, we note that in order to get meaningful results it may be useful to consider a Gaussian window rather than a top-hat which exhibits an extended tail at high wavenumbers.

\section{Conclusion}

Thus, in this article we have investigated the properties of the standard perturbative expansions which arise from the dynamics of gravitational clustering. First, we have shown that for hierarchical scenarios with no small-scale cutoff perturbative theory always breaks down beyond a finite order $q_+$ as divergent terms appear. Moreover, the degree of divergence increases with the order of the perturbative terms so that standard renormalization procedures do not apply. At first sight, the divergence of these subleading terms may cast some doubts on the predictions of perturbative theory obtained at lower orders. However, we have shown that the leading-order results can also be derived through a steepest-descent method which does not make use of these perturbative expansions. Therefore, the predictions obtained at leading-order for the cumulants of the density contrast are correct. This agrees with the results of numerical simulations which match these calculations (e.g., \cite{Ber2}). Then, we have shown that the simpler one-dimensional Burgers equation exhibits a similar behaviour. This analogy suggests that the results of perturbation theory are valid up to the order $q_+$ (i.e. until they are finite) and that beyond this order the large-scale expansion of the two-point correlation is no longer over powers of $\sigma^2(R)$ but over some different combination of powers of $1/R$.

In practice the linear power-spectrum shows a cutoff at a very small scale $R_c$, like for the CDM model (\cite{Peebles3}). This ensures that all integrals encountered in perturbative expansions are finite. However, it is clear that the results of perturbative theory break down when the contribution of non-linear scales becomes non-negligible. Thus, our analysis also applies to the cases where $R_c < R_0$, where $R_0$ is the scale which marks the transition to the non-linear regime. Then, to go beyond the order $q_+$ one needs again to devise new non-perturbative methods. On the other hand, if $R_c>R_0$ the standard perturbative expansions are sufficient, but this is not the case of cosmological interest. We can also note that although we considered a critical density universe so far, our analysis also extends to arbitrary cosmological parameters. 

Finally, we can point out that the leading-order terms are quite sufficient to describe the quasi-linear regime, up to $\sigma \la 1$, as seen in \cite{Ber2} and \cite{paper2}. Therefore, the break-up of perturbative expansions beyond a finite order is not a serious problem. Moreover, as discussed in \cite{paper1} (see also \cite{Bha1}) perturbative expansions do not capture any effect arising from multi-streaming. Hence, in order to derive convincing results in the early non-linear stages $\sigma \ga 1$ where a significant fraction of matter has experienced some shell-crossing (let us recall that a spherical density fluctuation collapses to a singularity as soon as $\delta_L \sim 1.68$) one should in any case use some non-perturbative approach. This still remains an open problem. However, we can note that some results in this direction have been obtained in \cite{paper4} for the rare-event tails of the probability distribution of the density contrast.

\end{document}